\documentclass[preprint,showpacs,preprintnumbers,amsmath,amssymb,onecolumn,superscriptaddress]{revtex4}

	\usepackage{graphicx,dcolumn,bm}

\begin{document}
\title{Mobilization of a trapped non-wetting fluid\\ from a three-dimensional porous medium}

    \author{Sujit S. Datta}
    \affiliation{Department of Physics, Harvard University, Cambridge MA 02138, USA}           
    \author{T. S. Ramakrishnan}
    \affiliation{Schlumberger-Doll Research, Cambridge, MA 02138, USA}

    \author{David A. Weitz}
    \affiliation{Department of Physics, Harvard University, Cambridge MA 02138, USA}
\email{weitz@seas.harvard.edu}

\date{\today}

\begin{abstract}   
We use confocal microscopy to directly visualize the formation and complex morphologies of trapped non-wetting fluid ganglia within a model 3D porous medium. The wetting fluid continues to flow around the ganglia after they form; this flow is characterized by a capillary number, $Ca$. We find that the ganglia configurations do not vary for small $Ca$; by contrast, as $Ca$ is increased above a threshold value, the largest ganglia start to become mobilized and are ultimately removed from the medium. By combining our 3D visualization with measurements of the bulk transport, we show that this behavior can be quantitatively understood by balancing the viscous forces exerted on the ganglia with the pore-scale capillary forces that keep them trapped within the medium. Our work thus helps elucidate the fluid dynamics underlying the mobilization of a trapped non-wetting fluid from a 3D porous medium. 
\end{abstract}
\maketitle

\section{Introduction}
Imbibition, the displacement of a non-wetting fluid from a porous medium by an immiscible wetting fluid, is crucial to many technological applications, including aquifer remediation, CO$_{2}$ sequestration, and oil recovery. When the three-dimensional (3D) pore space is highly disordered, the fluid displacement through it is complicated; this leads to the formation and trapping of discrete ganglia of the non-wetting fluid within the porous medium \cite{rose, herbeck,payatakes1, chatzis2, lenormand2,lenormand1,jerauld,blunt2,blunt1}. Some of this trapped non-wetting fluid becomes mobilized, and can ultimately become removed from the medium, as the capillary number characterizing the continued flow of the wetting fluid, $Ca$, is increased \cite{tabor,stegemeier,chatzis1,chatzis3,morrow2}. The pore-scale physics underlying this phenomenon remains intensely debated. Visual inspection of the exterior of a porous medium, as well as some simulations, suggest that as $Ca$ increases, the ganglia are not immediately removed from the medium; instead, they break up into smaller ganglia only one pore in size \cite{payatakes2,wardlaw1,2dmodel}. These then remain trapped within the medium, becoming mobilized and removed only for large $Ca$. By contrast, other simulations, as well as experiments on individual ganglia, suggest that the ganglia do not break up; instead, all ganglia larger than a threshold size, which decreases with increasing $Ca$, become mobilized and removed \cite{ng1,larson2,ng2,larson1,yortsos3}. The differences between these conflicting pictures, such as the geometrical configurations of the trapped ganglia, can have significant practical consequences: for example, smaller ganglia present a higher surface area per unit volume, potentially leading to their enhanced dissolution in the wetting fluid \cite{imhoff,chatzis4}. This behavior impacts diverse situations ranging from the spreading of contaminants in groundwater aquifers to the storage of CO$_{2}$ in brine-filled formations. Elucidating the physics underlying ganglion trapping and mobilization is thus critically important; however, despite its enormous industrial relevance, a clear understanding of this phenomenon remains lacking. Unfortunately, systematic experimental investigations of it are challenging, requiring direct measurements of the pore-scale ganglia configurations within a 3D porous medium, combined with measurements of the bulk transport through it, over a broad range of flow conditions. 

Here, we use confocal microscopy to directly visualize the formation and complex morphologies of the trapped non-wetting fluid ganglia within a model 3D porous medium. The ganglia vary widely in their sizes and shapes.  Intriguingly, these configurations do not vary for sufficiently small $Ca$; by contrast, as $Ca$ increases above a threshold value $\approx2\times10^{-4}$, the largest ganglia start to become mobilized and removed from the medium. Both the size of the largest trapped ganglion, and the total amount of trapped non-wetting fluid, decrease with $Ca$. We do not observe significant effects of ganglion breakup in our experiments. By combining our 3D visualization with measurements of the bulk transport properties of the medium, we show that the variation of the ganglia configurations with $Ca$ can instead be quantitatively understood using a mean-field model balancing the viscous forces exerted on the ganglia with the capillary forces that keep them trapped within the medium.

\section{Experimental methodology}
We prepare model 3D porous media by lightly sintering dense, disordered packings of hydrophilic glass beads, with radii $a = 19 \pm 1~\mu\mbox{m}$, in thin-walled square quartz capillaries of cross-sectional area $A = 9~\mbox{mm}^{2}$. The packings have length $L = 1~\mbox{cm}$; they thus span approximately 78 or 263 pores transverse to or along the imposed flow direction, respectively. Scattering of light from the surfaces of the beads typically precludes direct observation of the multiphase flow within the medium. We overcome this limitation by formulating Newtonian fluids whose compositions are carefully chosen to match their refractive indices with that of the glass beads \cite{amber}; the wetting fluid is a mixture of 91.4 wt\% dimethyl sulfoxide and 8.6 wt\% water, dyed with fluorescein, while the non-wetting oil is a mixture of aromatic and aliphatic hydrocarbons (immersion liquid, Cargille, 5040). The dynamic viscosities of the wetting fluid and the oil are $\mu_{w} = 2.7~\mbox{mPa s}$ and $\mu_{o} = 16.8~\mbox{mPa s}$, respectively, as measured using a stress-controlled rheometer. The densities of the wetting fluid and the oil are $\rho_{w} = 1.1~\mbox{g cm}^{-3}$ and $\rho_{o} = 0.83~\mbox{g cm}^{-3}$, respectively. The two fluids are fully immiscible over the experimental timescale; the interfacial tension between them is $\gamma = 13.0~\mbox{mN m}^{-1}$, as measured using a du No\"{u}y ring. The contact angle between the wetting fluid and glass in the presence of the oil is $\theta\approx 5^\circ$, as measured using confocal microscopy. 

Prior to each experiment, a porous medium is evacuated under vacuum and saturated with gaseous CO$_{2}$, which is soluble in the wetting fluid; this procedure eliminates the formation of trapped bubbles. We then saturate the pore space with the dyed wetting fluid; this enables us to visualize it in 3D using a confocal microscope, as schematized in Figure$~1$(a). We acquire 3D stacks of 39 optical slices parallel to the $xy$ plane, at $z$ positions at least several bead diameters deep within the medium; the slices each span an area of 912$~\mu\mbox{m}\times912~\mu\mbox{m}$ in the $xy$ plane, are 7$~\mu$m thick, and are spaced by $8~\mu\mbox{m}$ along the $z$ direction. We identify the glass beads by their contrast with the dyed wetting fluid. To visualize the pore structure of the entire medium, we acquire additional stacks, at the same $z$ positions, but at multiple $xy$ locations spanning the entire width and length of the medium. The porosity of the packings, measured using the 3D visualization, is $\phi=0.41\pm0.03$.

We subsequently flow 15 pore volumes of the non-wetting oil at a prescribed volumetric flow rate $Q_{o}=1~\mbox{mL h}^{-1}$ through the porous medium; this process is often referred to as primary drainage. Because the oil is undyed, we identify it by its additional contrast with the dyed wetting fluid in the imaged pore space. We then continuously flow the wetting fluid at a prescribed flow rate $Q_{w}$; this process is referred to as secondary imbibition. We characterize the wetting fluid flow using the capillary number $Ca\equiv\mu_{w}Q_{w}/A\gamma$; by progressively varying $Q_{w}$ over the range $10^{-1}-10^{3}~\mbox{mL h}^{-1}$, we explore four decades of $Ca\approx6\times10^{-7}-6\times10^{-3}$ in our experiments. The Reynolds number characterizing the pore-scale flow is given by $Re\equiv\rho_{w} (Q/\phi A)a_{t}/\mu_{w}\approx8\times10^{-6}-8\times10^{-2}$, where $a_{t}\approx0.16a$ is the typical radius of a pore entrance, and therefore, our experiments are characterized by slow, viscous flow. The Bond number characterizing the influence of gravity relative to capillary forces at the pore scale is given by $Bo\equiv g(\rho_{w}-\rho_{o})a_{t}^{2}/\gamma\approx10^{-6}$, indicating that gravity likely influences trapping only above vertical length scales comparable to that of the entire medium. We therefore neglect gravity from our subsequent theoretical analysis.

To visualize the dynamics of the fluid displacement during both primary drainage and secondary imbibition, we repeatedly acquire a series of optical slices, at a fixed $z$ position several bead diameters deep within the medium, but at multiple $xy$ locations spanning the entire length of the medium. This procedure thus enables us to directly visualize the multiphase flow, both at the scale of the individual pores and at the scale of the overall medium.

The secondary imbibition leads to the formation of discrete oil ganglia, many of which remain trapped within the pore space. To probe the ganglia configurations, we reacquire a second set of 3D stacks at the same $x$, $y$, and $z$ positions as the stacks obtained during the initial characterization of the pore structure. By comparing the two sets of stacks, we obtain the 3D morphologies of the trapped oil ganglia at sub-pore resolution. We restrict our analysis to an area several beads away from each edge of the medium to minimize boundary effects. To explore the variation of the ganglia configurations with the imposed flow conditions, we increase $Q_{w}$ in increments. For each value of $Q_{w}$, we flow $>13$ pore volumes of the wetting fluid, thus establishing a new steady state, before reacquiring an additional set of 3D stacks. By again comparing each set of stacks with that obtained during the initial pore structure characterization, we obtain the 3D morphologies of the oil ganglia left trapped at each $Ca$.

To quantify the bulk transport behavior, we use differential pressure sensors to measure the pressure drop $\Delta P$ across a porous medium constructed in a manner similar to that used for the 3D visualization. We first saturate the medium with the wetting fluid and vary $Q_{w}$; by measuring the proportionate variation in $\Delta P$, we determine the single-phase permeability of the medium, $k\equiv\mu_{w}(Q_{w}L/A)/\Delta P=1.67~\mu\mbox{m}^{2}$. The permeability of a disordered packing of monodisperse spheres is typically estimated using the Kozeny-Carman relation, $k=\frac{1}{45}\frac{\phi^{3}a^{2}}{(1-\phi)^{2}}$ \cite{philipse}; this yields $k=1.59~\mu\mbox{m}^{2}$, in excellent agreement with our measured value. We then flow the oil, and reflow the wetting fluid, through the porous medium, following the same procedure as that used for the 3D visualization. The oil trapped after secondary imbibition occludes some of the pore space, modifying the flow through it; this reduces the wetting fluid permeability to a value $\kappa k$, where $\kappa\leq1$ is known as the end-point relative permeability. For each $Ca$ investigated, we flow $>13$ pore volumes of the wetting fluid at its corresponding $Q_{w}$, and then measure $\Delta P$ at $Q_{w}=10^{-1}~\mbox{mL h}^{-1}$. Comparing $\Delta P$ with that measured during single-phase flow at $Q_{w}=10^{-1}~\mbox{mL h}^{-1}$ directly yields $\kappa$.
 \begin{figure}
  \centerline{\includegraphics[width=13.5cm]{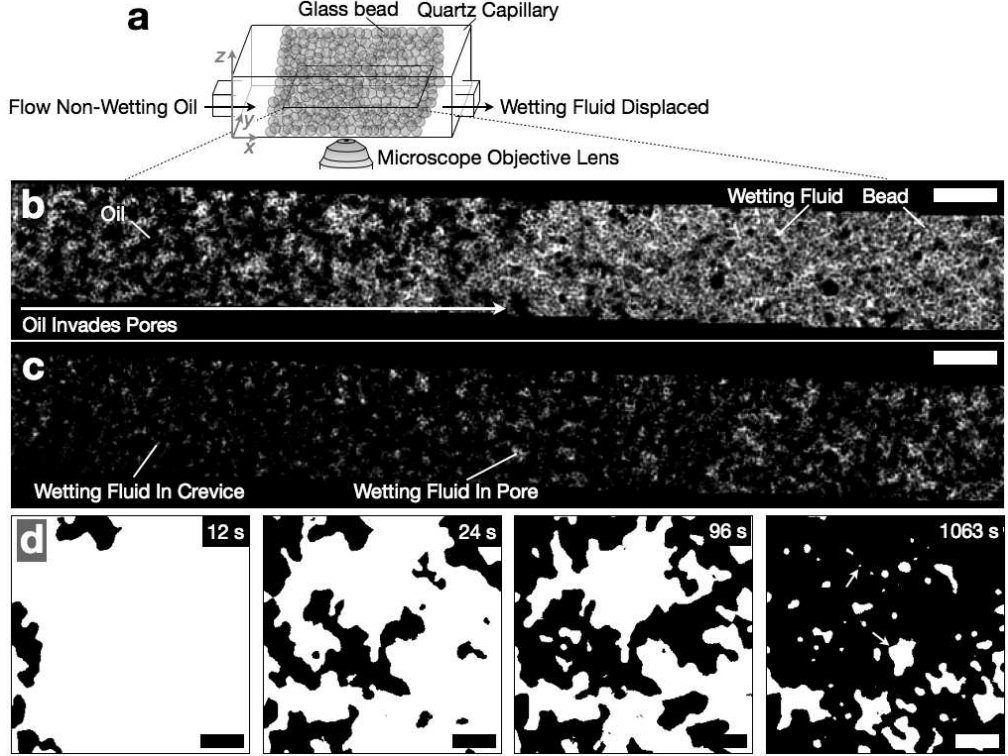}}
  \caption{(a) Schematic of primary drainage through a model 3D porous medium. We directly visualize the flow within the medium using confocal microscopy. The pore space is initially saturated with the fluorescently-dyed wetting fluid, which is displaced by the undyed non-wetting oil.   (b) Optical section through part of the medium, taken as the oil displaces the wetting fluid at $Q_{o}=1~\mbox{mL h}^{-1}$. Section is obtained at a fixed $z$ position, away from the lateral boundaries of the medium. Bright areas show the pore space, saturated with the fluorescently-dyed wetting fluid, and the black circles show cross-sections of the beads making up the medium. Additional black areas show the invading oil. The path taken by the oil varies spatially, as seen in the region spanned by the arrow. (c) Optical section through the same part of the medium, taken after invasion by $\approx 9$ pore volumes of the oil. Some wetting fluid remains trapped in the crevices and pores of the medium, as indicated. (d) Time sequence of zoomed confocal micrographs, with the pore space subtracted; binary images thus show oil in black as it bursts into the pores. Time stamp indicates time elapsed after subtracted frame. Upper and lower arrows in the last frame show wetting fluid trapped in a crevice or in a pore, respectively. Scale bars in (b-c) and (d) are $500~\mu\mbox{m}$ and $200~\mu\mbox{m}$, respectively. Imposed flow direction in all images is from left to right.}
\end{figure}

\begin{figure}
  \centerline{\includegraphics[width=13.5cm]{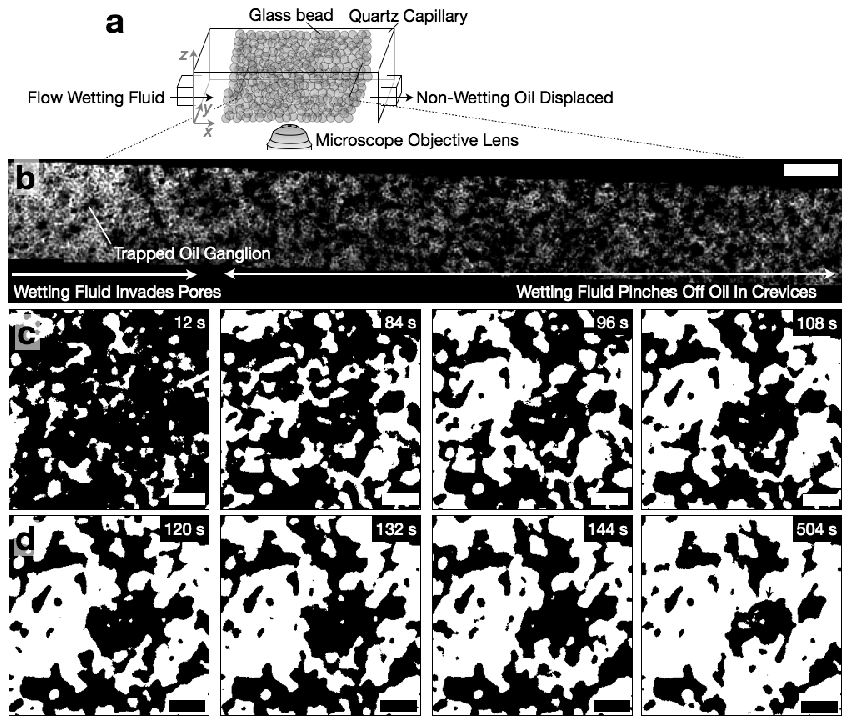}}
  \caption{(a) Schematic of secondary imbibition through a model 3D porous medium. We directly visualize the flow within the medium using confocal microscopy. The fluorescently-dyed wetting fluid displaces the undyed non-wetting oil. (b) Optical section through part of the medium, taken as the wetting fluid displaces the oil at $Ca=6.4\times10^{-7}$. Section is obtained at the same fixed $z$ position, away from the lateral boundaries of the medium, as in Figure$~1$(b-c). Bright areas show the fluorescently-dyed wetting fluid, and the black circles show cross-sections of the beads making up the medium. Additional black areas show the oil. The wetting fluid first snaps off oil in crevices throughout the medium, as seen in the region spanned by the double-headed arrow, and then bursts into the pores of the medium, starting at the inlet, as seen in the region spanned by the single-headed arrow. Some oil ganglia remain trapped within the medium, as indicated.  (c-d) Time sequence of zoomed confocal micrographs, with the oil-filled pore space subtracted; binary images thus show wetting fluid in white as it (c) initially snaps off oil in the crevices, and then (d) invades the pores. Time stamps indicate time elapsed after subtracted frame. Last frame shows unchanging steady state; arrow indicates a trapped oil ganglion. Scale bars in (b) and (c-d) are $500~\mu\mbox{m}$ and $200~\mu\mbox{m}$, respectively. Imposed flow direction in all images is from left to right.}
\end{figure}

\section{Results and discussion}
\subsection{Dynamics of multiphase flow}
To mimic the migration of a non-wetting fluid into a geological formation, we first drain the 3D porous medium, initially saturated with the dyed wetting fluid, with the undyed non-wetting oil at $Q_{o}=1~\mbox{mL h}^{-1}$. The oil displaces the wetting fluid through a series of intermittent, abrupt bursts into the pores; this indicates that a threshold capillary pressure difference must build up in the oil before it can invade a pore \cite{mohanty1,datta1}. This pressure is given by $2\gamma\cos\theta/a_{t}$, where $a_{t}\approx0.16a$ is the typical radius of a pore entrance \cite{lenormand2,mayer,princen1,princen2,princen3,mason1,toledo1,raoush,thompson}. Because the packing of the beads is disordered, $a_{t}$ varies from pore to pore, forcing the path taken by the invading oil to similarly vary spatially, as exemplified by the optical sections in Figures$~1$(b) and (d) \cite{lenormand3,martys1,maloy1,xu1,xu2}. Consequently, the interface between the oil and the displaced wetting fluid interface is ramified [tip of the arrow in Figure$~1$(b)]. As the oil continues to drain the medium, it eventually fills most of the pore space, as shown in Figure$~1$(c); however, the smallest pores remain filled with the wetting fluid [rightmost indicator in Figure$~1$(c) and lower arrow in Figure$~1$(d)]. Thin layers of the wetting fluid, $\approx1~\mu\mbox{m}$ thick, remain trapped in the crevices of the medium, surrounding the oil in the pores [leftmost indicator in Figure$~1$(c) and upper arrow in Figure$~1$(d)]. This observation provides direct confirmation of the predictions of a number of theoretical calculations and numerical simulations \cite{lenormand1,ramaflow,diaz,blunt5,blunt2,blunt4,blunt3,mogensen,oren,constantinides,hughes,patzek,blunt1,idowu}.

To investigate the dynamics of secondary imbibition, we then flow the wetting fluid continuously at $Ca=6.4\times10^{-7}$. The presence of the thin layers of the wetting fluid profoundly changes the flow dynamics: unlike the case of primary drainage, the invading fluid does not simply burst into the pores. Instead, we observe that the wetting fluid first flows through the thin layers, snapping off the oil in crevices throughout the entire medium \cite{mohanty2,yu,roof} over a period of approximately $60~\mbox{s}$, as seen in the region spanned by the double-headed arrow in Figure$~2$(b) and in the time sequence shown in Figure$~2$(c). This behavior agrees with the predictions of recent numerical simulations \cite{nguyen}. The wetting fluid then also begins to invade the pores through a series of intermittent, abrupt bursts, starting from the inlet, as seen in the region spanned by the single-headed arrow in Figure$~2$(b) and in the time sequence shown in Figure$~2$(d). Interestingly, as the wetting fluid invades the medium, it bypasses many of the pores, leaving discrete oil ganglia of varying sizes in its wake. Many of these ganglia remain trapped within the pore space, as indicated in Figure$~2$(b).

\subsection{Trapped oil ganglia configurations}
We use our confocal micrographs to measure the total amount of oil trapped within the porous medium; this is quantified by the residual oil saturation, $S_{or}\equiv V_{o}/\phi V$, where $V_{o}$ is the total volume of oil imaged within a region of volume $V$. After secondary imbibition at $Ca=6.4\times10^{-7}$, we find $S_{or}\approx9\%$, similar to the results of some previous experiments on bead packs \cite{morrow2}. To mimic discontinuous core-flood experiments on reservoir rocks, we then explore the variation of $S_{or}$ in response to progressive increases in the wetting fluid $Ca$. We find that $S_{or}$ does not vary significantly for sufficiently small $Ca$; however, as $Ca$ is increased above $2\times10^{-4}$, $S_{or}$ decreases precipitously, as shown in Figure$~3$, ultimately reaching only $\approx7\%$ of its initial value. These results are consistent with the results of previous core-flood experiments, which show similar behavior for fluids of a broad range of viscosities and interfacial tensions \cite{morrow2}. The measured threshold value of $Ca$ is also consistent with the results of some numerical simulations \cite{yortsos3}.

\begin{figure}
  \centerline{\includegraphics[width=11cm]{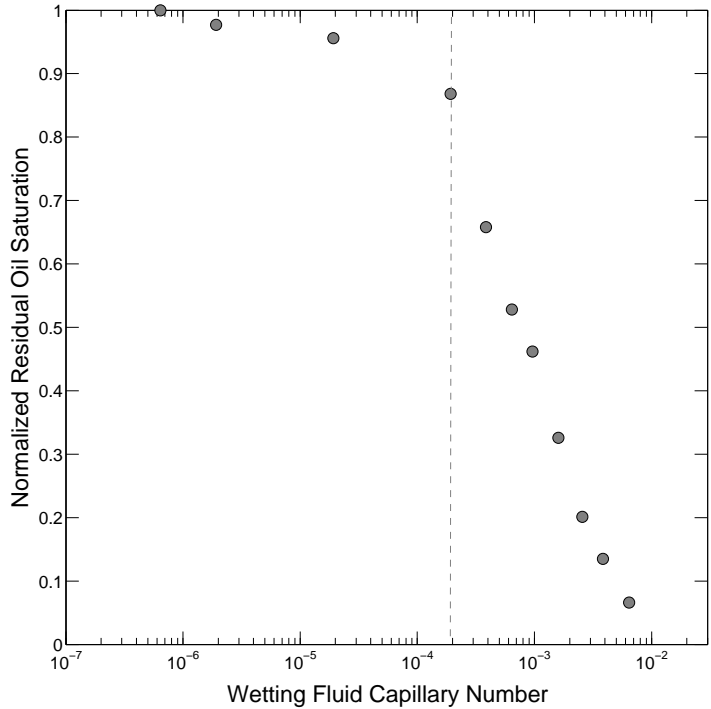}}
  \caption{Residual oil saturation $S_{or}$, normalized by its maximum value, does not vary significantly for small wetting fluid capillary number $Ca$, but decreases precipitously as $Ca$ increases above $2\times10^{-4}$ (dashed grey line). Residual oil saturation is measured using 3D confocal micrographs.}
\end{figure}

\begin{figure}
  \centerline{\includegraphics[width=9cm]{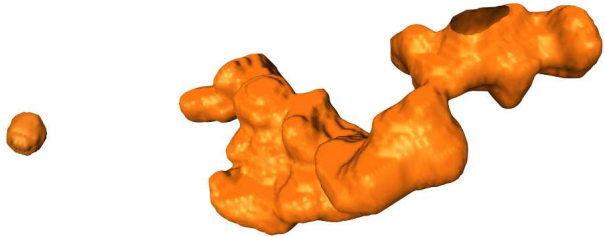}}
  \caption{3D renderings of oil ganglia trapped after secondary imbibition at $Ca=6.4\times10^{-7}$. The left ganglion is spherical and only spans $\approx0.3$ beads in diameter, whereas the right ganglion is more ramified and spans multiple beads. Renderings are produced using 3D confocal micrographs.}
\end{figure}

\begin{figure}
  \centerline{\includegraphics[width=11cm]{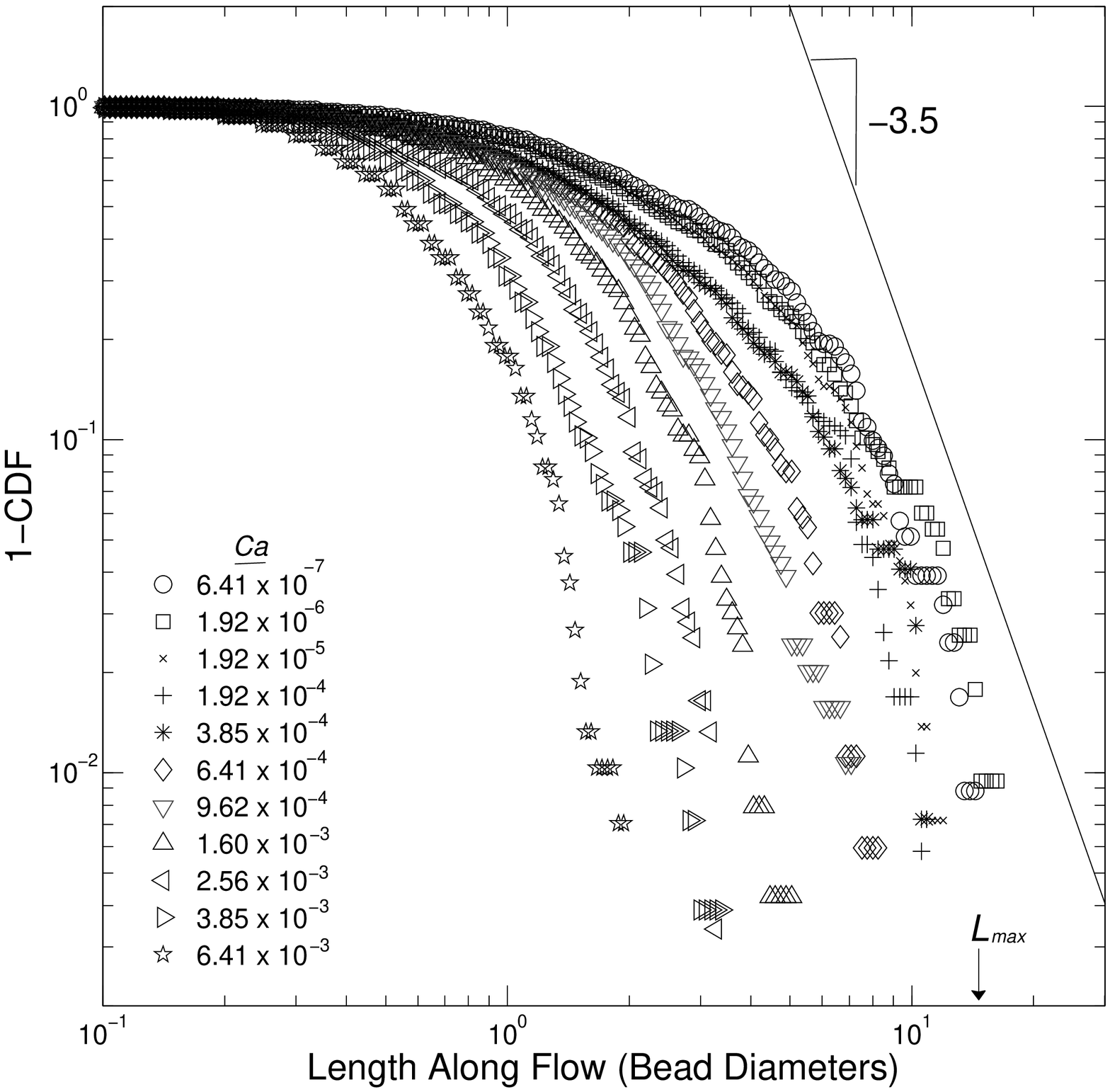}}
  \caption{Complementary cumulative distribution functions 1-CDF of ganglia lengths $L$, measured along the flow direction using 3D confocal micrographs, after secondary imbibition at a range of wetting fluid capillary numbers $Ca$.  Solid line shows $\sim L^{-3.5}$ scaling. Arrow indicates maximum ganglion length $L_{max}$ at the lowest $Ca=6.41\times10^{-7}$.}
\end{figure}

To better understand this behavior, we inspect the reconstructed 3D morphologies of the individual ganglia for each $Ca$ investigated. At the smallest $Ca\approx6\times10^{-7}-2\times10^{-4}$, the ganglia morphologies vary widely, as exemplified by the 3D renderings shown in Figure$~4$. The smallest ganglia are spherical, only occupying single pores, and span $\approx0.3$ beads in size [left, Figure$~4$]; in stark contrast, the largest ganglia are ramified, occupying multiple pores, and span many beads in size [right, Figure$~4$]. To quantify the significant variation in their morphologies, we measure the length $L$ of each ganglion along the flow direction, and plot $\mbox{1-CDF}(L)$, where $\mbox{CDF}=\sum_{0}^{L}Lp(L)/\sum_{0}^{\infty}Lp(L)$ is the cumulative distribution function of ganglia lengths and $p(L)$ is the number fraction of ganglia having a length $L$. Consistent with the variability apparent in the 3D renderings, we find that the ganglia lengths are broadly distributed, as indicated by the circles in Figure$~5$. Interestingly, in agreement with recent X-ray microtomography experiments \cite{iglauer1,iglauer2,berg}, we find that the decay of this distribution is consistent with a power law, $\mbox{1-CDF}(L)\propto L^{-3.5}$ [solid line in Figure$~5$], as predicted by percolation theory [Appendix 1]. Moreover, we find that the largest trapped ganglion has a length $L_{max}\approx13$ bead diameters [arrow in Figure$~5$]; while we cannot exclude the influence of boundary effects or the limited imaging volume, this value is in good agreement with the prediction of percolation theory incorporating a non-zero viscous pressure: $L_{max}\approx\alpha(a^{2}Ca/\kappa k)^{-\nu/(1+\nu)}\approx9\alpha$ bead diameters, where $\nu\approx0.88$ is a scaling exponent obtained from numerical simulations \cite{wilkinson2}, $\alpha$ is a constant of order unity, and we use the value of $\kappa$ measured at the lowest $Ca$ \cite{wilkinson2,yortsosnote}. Taken together, these results suggest that the configurations of the ganglia left trapped after secondary imbibition are consistent with the predictions of percolation theory. 

\begin{figure}
  \centerline{\includegraphics[width=11cm]{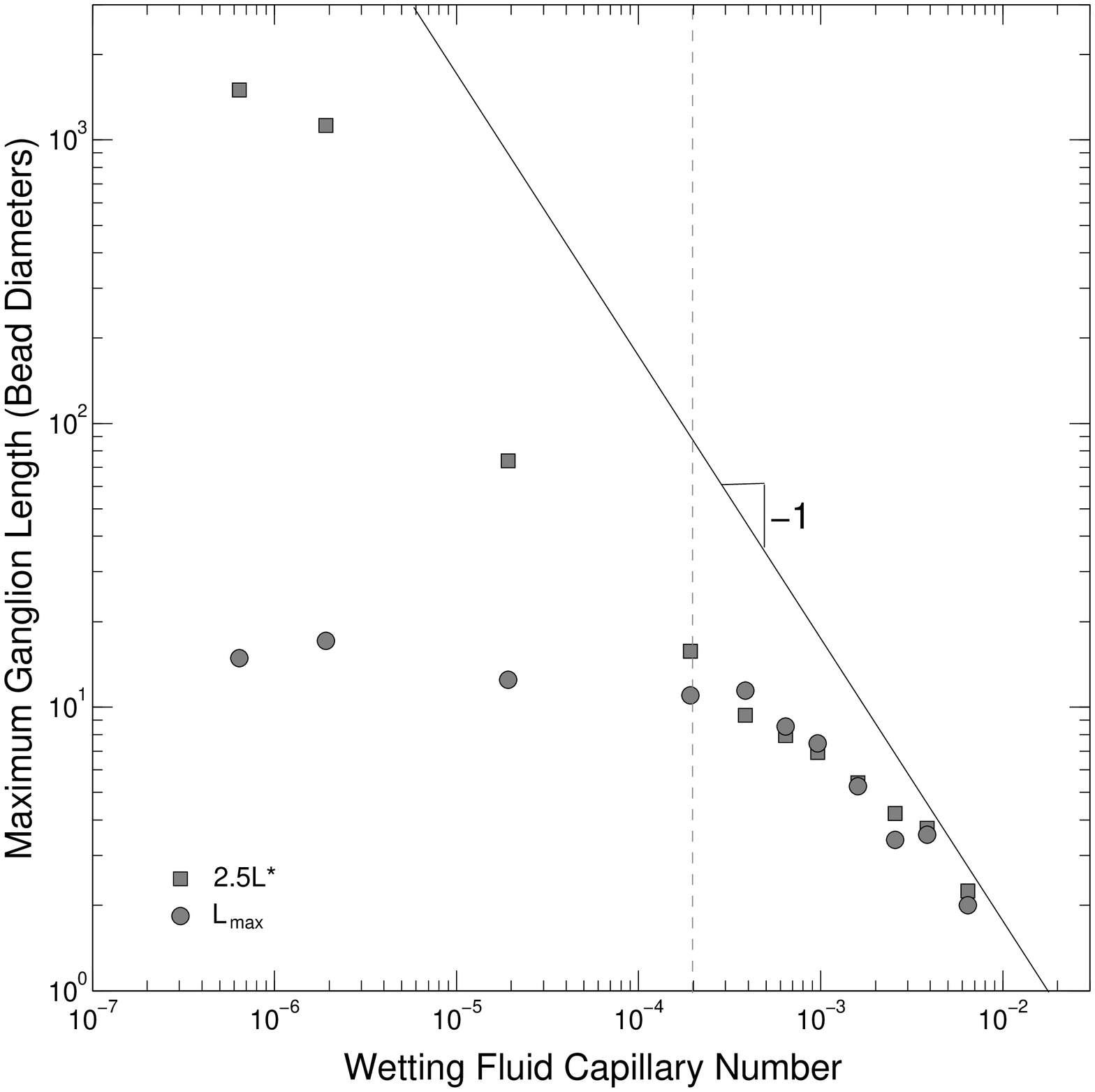}}
  \caption{Variation of maximum trapped oil ganglion length $L_{max}$, measured along the flow direction using 3D confocal micrographs, and 2.5 times the theoretical $L^{*}$, calculated using Eq. (3.3), with the wetting fluid capillary number $Ca$. Similar to the variation of the residual oil saturation $S_{or}$, shown in Figure$~3$, $L_{max}$ does not vary significantly for small wetting fluid capillary number $Ca$, but decreases precipitously as $Ca$ increases above $2\times10^{-4}$ (dashed grey line). Both $L_{max}$ and $L^{*}$ do not scale as $\sim Ca^{-1}$, indicated by the solid line.}
\end{figure}

As $Ca$ increases, we do {\it not} observe significant effects of ganglia breakup; this observation is contrary to some previous suggestions \cite{wardlaw1}, and confirms the predictions of other numerical simulations \cite{ng2,breakupnote}. Instead, the ganglia configurations remain the same for all $Ca<2\times10^{-4}$ [$\circ$, $\square$, and $\times$ in Figure$~5$]. Moreover, we find that the largest ganglia start to become mobilized and removed from the porous medium, concomitant with the observed decrease in $S_{or}$, once $Ca$ increases above $2\times10^{-4}$ [Figure$~5$]. We quantify this behavior by plotting the variation of $L_{max}$ with $Ca$. While $L_{max}$ remains constant at small $Ca$, it decreases precipitously as $Ca$ increases above $2\times10^{-4}$, as shown by the circles in Figure$~6$. Remarkably, this behavior closely mimics the observed variation of $S_{or}$ with $Ca$ [Figure$~3$]. These results thus suggest that the variation of $S_{or}$ with increasing $Ca$ is not determined by the breakup of the trapped ganglia; instead, it may reflect the mobilization and removal of the largest ganglia from the medium. 

\begin{figure}
  \centerline{\includegraphics[width=11cm]{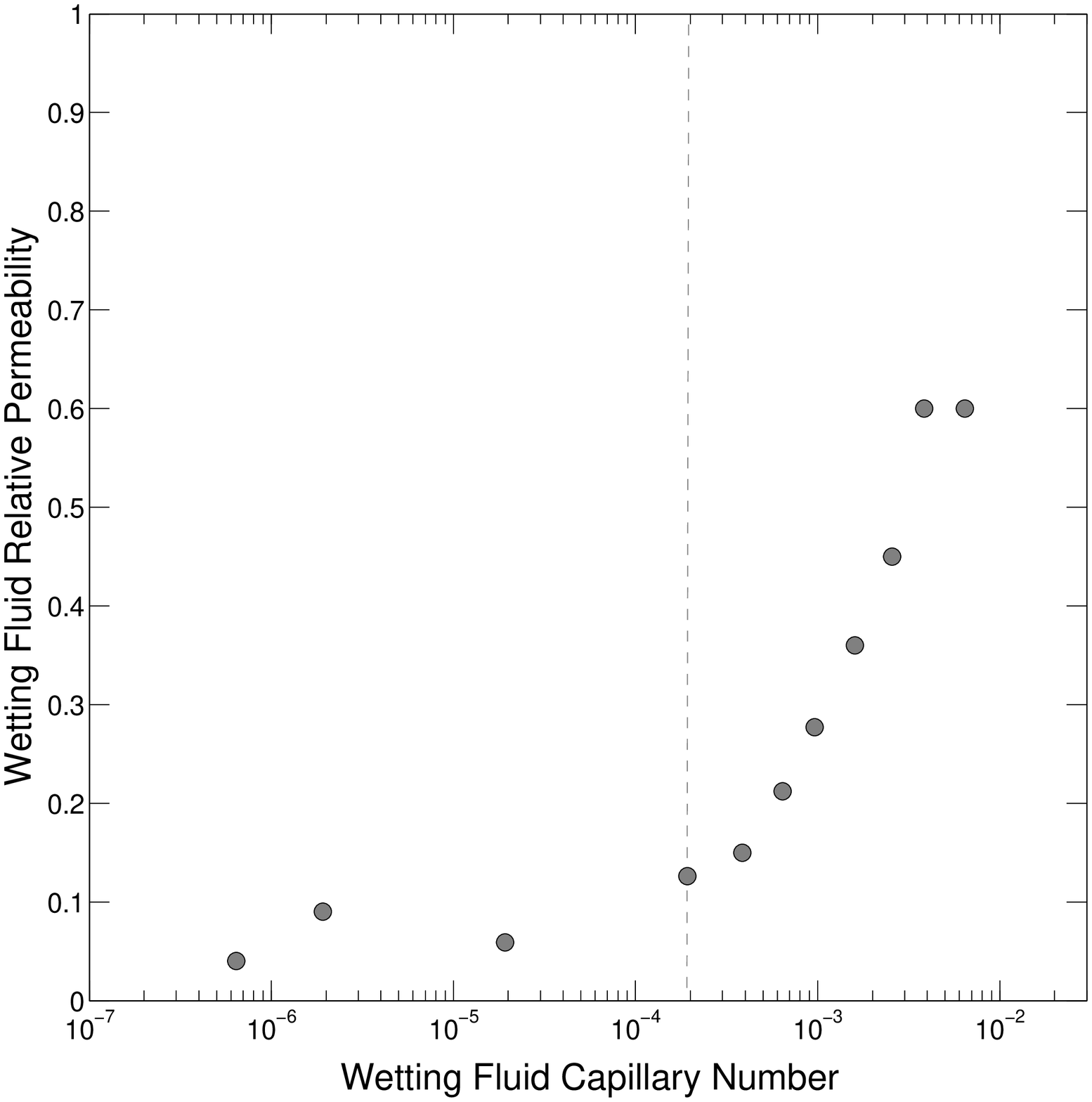}}
  \caption{Variation of wetting fluid end-point relative permeability $\kappa$, measured using pressure transducers, with the wetting fluid capillary number $Ca$. Similar to the variation of the residual oil saturation $S_{or}$, shown in Figure$~3$, $\kappa$ does not vary significantly for small wetting fluid capillary number $Ca$; however, it increases dramatically as $Ca$ increases above $2\times10^{-4}$ (dashed grey line).}
\end{figure}

\subsection{Physics of ganglion mobilization}
To test this hypothesis, we analyze the distribution of pressures in the wetting fluid as it flows through the porous medium. Motivated by previous studies of this flow \cite{gardesu,ng1,chatzis1,datta2}, we make the mean-field assumption that the viscous pressure drop across a ganglion of length $L$ is given by Darcy's law,
\begin{equation}
P_{v}=\frac{\mu_{w}}{\kappa k}\frac{Q_{w}}{A}L
\end{equation}
The end-point relative permeability $\kappa\leq1$ quantifies the modified transport through the medium due to the presence of the trapped oil. To determine $P_{v}$ at each $Ca$ investigated, we directly measure the variation of $\kappa$ with $Ca$ for a porous medium constructed in a manner similar to, and following the same flow procedure as, that used for visualization of the ganglia configurations. Interestingly, $\kappa$ does not vary significantly for sufficiently small $Ca$; however, as $Ca$ increases above $2\times10^{-4}$, $\kappa$ quickly increases, concomitant with the observed decreased in $S_{or}$, as shown in Figure$~7$. This observation suggests that the bulk transport behavior of the medium depends strongly on the trapping of oil within it. To quantify the close link between the variation of $\kappa$ and $S_{or}$ with $Ca$ \cite{morrow2}, we plot $\kappa$ as a function of the wetting fluid saturation, $1-S_{or}$. Consistent with our expectation \cite{powder}, we find that $\kappa$ increases monotonically with increasing wetting fluid saturation, as shown in Figure$~8$. 

\begin{figure}
  \centerline{\includegraphics[width=11cm]{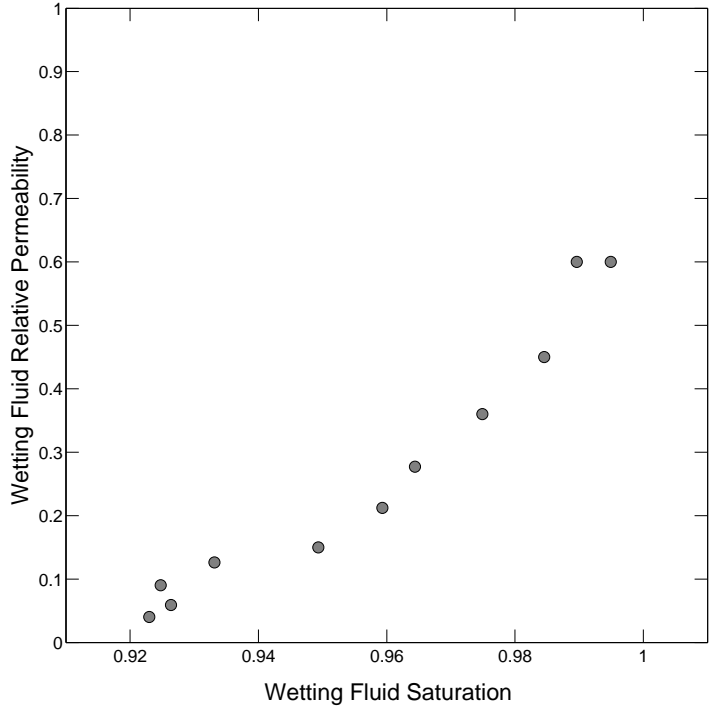}}
  \caption{Variation of wetting fluid end-point relative permeability $\kappa$, measured using pressure transducers, with the saturation of wetting fluid in the pore space, $1-S_{or}$. We find that $\kappa$ increases monotonically as the amount of trapped oil is decreased.}
\end{figure}

For a ganglion to squeeze through the pores of the medium, it must simultaneously displace the wetting fluid from a downstream pore, and be displaced by the wetting fluid from an upstream pore. To displace the wetting fluid from a downstream pore, a threshold capillary pressure must build up at the pore entrance, as schematized by the right set of arrows in Figure$~9$; this threshold is given by $P_{t}=2\gamma\cos\theta/a_{t}$, where $a_{t}$ is the radius of the pore entrance \cite{lenormand1,mayer,princen1,princen2,princen3,mason1,toledo1}, with an average value $\approx0.16a$ for a 3D packing of glass beads \cite{raoush,thompson}. Similarly, for the trapped oil to be displaced from an upstream pore, the capillary pressure within the pore must fall below a threshold, as schematized by the left set of arrows in Figure$~9$; this threshold is given by  $P_{b}=2\gamma\cos\theta/a_{b}$, where $a_{b}$ is instead the radius of the pore itself, with an average value $\approx0.24a$ \cite{thompson,raoush}. Thus, to mobilize a ganglion and ultimately remove it from the porous medium, the total viscous pressure drop across it must exceed a capillary pressure threshold, 
\begin{equation}
P_{t}-P_{b}=\frac{2\gamma\cos\theta}{a_{b}}\left(\frac{a_{b}}{a_{t}}-1\right)
\end{equation} 
Balancing Eqs. (1) and (2), we therefore expect that \cite{frictionnote}, at a given $Ca$, the smallest ganglia remain trapped within the medium, while the viscous pressure in the wetting fluid is sufficiently large to mobilize all ganglia larger than
\begin{equation}
L_{max}= L^{*}\equiv\frac{2\cos\theta}{Ca}\left(\frac{a_{b}}{a_{t}}-1\right)\frac{\kappa k}{a_{b}}
\end{equation}
To critically test this prediction, we compare the variation of both $L_{max}$, directly measured using confocal microscopy, and $L^{*}$, calculated using the measured values of $\theta$, $k$, and $\kappa$, with $Ca$. For small $Ca$, we find $L_{max}<L^{*}$, as shown by the first three points in Figure$~6$; this indicates that the viscous pressure in the wetting fluid is too small to mobilize any ganglia. Consequently, $S_{or}$ does not vary significantly for this range of $Ca$, consistent with the measurements shown in Figure 3. As $Ca$ increases, $L_{max}$ remains constant; however, $L^{*}$ steadily decreases, eventually becoming comparable to $L_{max}$ at $Ca\approx2\times10^{-4}$, shown by the dashed line in Figure$~6$. Strikingly, as $Ca$ increases above this value, we find that both $L_{max}$ and $L^{*}$ decrease in a similar manner, with $L_{max}\approx 2.5 L^{*}$; this indicates that the viscous pressure in the wetting fluid is sufficient to mobilize and remove more and more of the largest ganglia. Consequently, we expect $S_{or}$ to also decrease with $Ca$ in this range, in excellent agreement with our measurements [Figure$~3$]. The similarity in the variation of $L_{max}$ and $S_{or}$ with $Ca$, and the close agreement between our measured $L_{max}$ and the predicted $L^{*}$ for $Ca>2\times10^{-4}$, thus confirm that the reduction in $S_{or}$ reflects the mobilization of the largest ganglia. Finally, we note that, due to the concomitant variation of the wetting fluid permeability with $Ca$, the measured $L_{max}$ does {\it not} decrease as $Ca^{-1}$ [solid line in Figure$~6$]; this is in contradiction to results obtained for an isolated ganglion \cite{ng1}, for which the permeability is a constant, which are often assumed to also apply to a population of many ganglia.

\begin{figure}
  \centerline{\includegraphics[width=8cm]{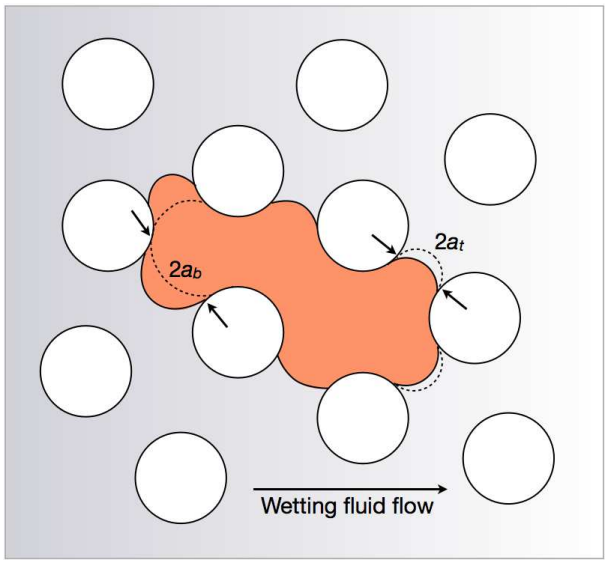}}
  \caption{2D schematic showing an oil ganglion (orange) trapped within the pore space, with wetting fluid flowing from left to right; beads are shown by white circles. Dashed lines show threshold curvatures required for the ganglion to invade the downstream pores or be displaced from the upstream pores.}
\end{figure}

\section{Conclusions}
Using confocal microscopy, we directly visualize the dynamics of primary drainage and secondary imbibition, as well as the intricate morphologies of the resultant trapped non-wetting fluid ganglia, within a 3D porous medium, at pore-scale resolution. During imbibition, the wetting fluid first flows through thin layers coating the solid surfaces, snapping off the non-wetting fluid in crevices throughout the medium. It then displaces the non-wetting fluid from the pores of the medium through a series of intermittent, abrupt bursts, starting from the inlet, leaving ganglia of the non-wetting fluid in its wake. These vary widely in their sizes and shapes, consistent with the predictions of percolation theory. We do not observe significant effects of ganglion breakup, contrary to some previous suggestions. Instead, for small $Ca$, the ganglia configurations do not appreciably change; this likely reflects the fact that the ganglia sizes, and hence the viscous pressure drops across them, are limited, consistent with the predictions of percolation theory incorporating a non-zero viscous pressure gradient. However, as $Ca$ is increased above a threshold value, more and more of the largest ganglia become mobilized and removed from the medium. By coupling the 3D visualization and bulk transport measurement, we show that the variation of the ganglia configurations can be understood by balancing the viscous forces exerted on the ganglia with the pore-scale capillary forces that keep them trapped within the medium. This work thus helps elucidate the fluid dynamics underlying the mobilization of a trapped non-wetting fluid from a 3D porous medium.

Our results provide direct visualization of the multiphase flow and the ganglia configurations within a 3D porous medium; moreover, they highlight the applicability of mean-field ideas in understanding the mobilization of the trapped non-wetting fluid. This work may thus help guide theoretical models or numerical simulations (e.g. \cite{rama,tyagi}). Moreover, because many geophysical flows give rise to residual trapping, we expect that our work will be relevant to a number of important applications, including enhancing oil recovery, understanding the distribution of contaminants in groundwater aquifers, or the storage of CO$_{2}$ in sub-surface formations. 

\acknowledgements 
It is a pleasure to acknowledge D. L. Johnson and J. R. Rice for stimulating discussions, and S. Khoulani for experimental assistance. This work was supported by the NSF (DMR-1006546), the Harvard MRSEC (DMR-0820484), and the Advanced Energy Consortium (http://www.beg.utexas.edu/aec/), whose member companies include BP America Inc., BG Group, Petrobras, Schlumberger, Shell, and Total. SSD acknowledges funding from ConocoPhillips. 

\section{Appendix}
Percolation theory predicts that the number fraction of ganglia with volume $s$ is, for large $s$, given by $p(s)\propto s^{-\tau}$, where $\tau\approx2.2$ is a scaling exponent \cite{gaunt,hoshen,stauffer}. Moreover, the volume of a ganglion is related to its length by the relation $s\propto L^{3/(\tau-1)}$. Combining these two relations yields $p(L)\propto L^{-3\tau/(\tau-1)}$, and therefore, $Lp(L)\propto L^{-3\tau/(\tau-1)+1}$. In our experiments, we measure the complementary cumulative distribution function of ganglia lengths, $\mbox{1-CDF}(L)=\sum_{L}^{\infty}Lp(L)/\sum_{0}^{\infty}Lp(L)\propto L^{-3\tau/(\tau-1)+2}\propto L^{-3.5}$ for $\tau\approx2.2$. This prediction is in good agreement with the large $L$ tail of our data, as shown by the solid line in Figure 5.


\begin{thebibliography}{62}
\bibitem{yortsosnote}
We note that this theoretical prediction does not incorporate the effects of ganglion breakup, consistent with our experimental observations; it also does not explicitly incorporate the effects of interactions between the different ganglia.
\bibitem{breakupnote}
It is unclear why we do not observe significant ganglion breakup in our experiments. This process may depend sensitively on the geometry of the pore space; for example, it may be appreciable in more disordered pore spaces that are characterized by larger pore body/pore throat aspect ratios.
\bibitem{frictionnote}
For simplicity, we only consider viscous and capillary forces. Including other factors into this picture would be a useful direction for future work.
\end{thebibliography}
\end{document}